\documentclass[12pt]{article}

\usepackage{newtxtext,newtxmath}
\usepackage{graphicx}
\usepackage[letterpaper,margin=1in]{geometry}
\usepackage[normalem]{ulem}

\renewenvironment{abstract}
  {\quotation}
  {\endquotation}

\date{}

\makeatletter
\renewcommand{\fnum@figure}{\textbf{Figure \thefigure}}
\renewcommand{\fnum@table}{\textbf{Table \thetable}}
\makeatother

\usepackage{scicite}
\usepackage{url}

\def\scititle{
  On-chip Quantum Measurement of Squeezing Generated from a Silicon Nitride Micro-ring Resonator
}
\title{\bfseries \boldmath \scititle}

\author{
  Yuhang~Lei$^{1\dagger}$,
  Chenfei~Cui$^{2\dagger}$,
  Yue~Li$^{1,3}$,
  Hon~Ki~Tsang$^{2\ast}$,
  Z.~Y.~Ou$^{1\ast}$\and
  \small$^{1}$Department of Physics, City University of Hong Kong,
  Kowloon, Hong Kong SAR, P. R. China.\and
  \small$^{2}$Department of Electronic Engineering,
  The Chinese University of Hong Kong, Shatin,
  Hong Kong SAR, P. R. China.\and
  \small$^{3}$Department of Electrical Engineering,
  City University of Hong Kong, Kowloon,
  Hong Kong SAR, P. R. China.\and
  \small$^\ast$Corresponding authors. Email:
  hktsang@ee.cuhk.edu.hk (H.K.T.);
  jeffou@cityu.edu.hk (Z.Y.O.)\and
  \small$^\dagger$These authors contributed equally to this work.
}

\begin{document}

\maketitle

\begin{abstract} \bfseries \boldmath
Integration of quantum optical technique on-chip is crucial for large scale applications of quantum technology, which were proven in a free space environment to be superior to the corresponding classical technology. Squeezed states of light can be used for enhancing the sensitivity of quantum sensors and for fault-tolerant quantum computing. Although chip-based squeezed light generation has advanced significantly, practical impact remains limited because coupling losses between the chip and off-chip detectors destroy delicate quantum correlations, restricting the amount of observed squeezing. Here, we overcome this limitation by implementing the idea of on-chip quantum measurement with the aid of a parametric amplifier and applying it to the squeezed state generated by a silicon nitride (SiN) micro-ring resonator. In our scheme, two matched SiN micro-rings are sequentially constructed. The first ring generates a squeezed state, whereas the second ring acts as a high-gain parametric amplifier (PA) that measures the squeezed state before the light experiences significant off-chip loss. This architecture is inherently loss-tolerant: the amplifier elevates the quantum noise well above the vacuum level, making the measurement insensitive to downstream losses. We directly observe a quantum noise reduction of 4.6 dB from the first ring, despite a chip-to-fiber coupling loss exceeding 5 dB. This work also demonstrates the first monolithic SU(1,1) interferometer with an estimated 5 dB signal-to-noise enhancement compared to traditional linear interferometers, and thus establishes a practical pathway for chip-based quantum sensors.

\end{abstract}

\section*{Introduction}
\noindent The drive to miniaturize quantum optical systems from table-top configurations to monolithic photonic integrated circuits is central to the development of scalable optical quantum technologies for large scale applications. Significant progress has been made in the generation of squeezed light, a key resource for quantum sensing and quantum computing, using on-chip micro-ring resonators and nonlinear waveguides \cite{dutt2015chip,chen2022ultra,zhao2020near,vaidya2020broadband,zhang2021squeezed,yang2021squeezed}. However, the impact of these sources has been stifled by a fundamental practical issue: after the squeezed light is generated on-chip, it must be coupled out to off-chip detectors. The accompanying optical losses in chip coupling (typically $>$ 4 dB per facet) rapidly destroy fragile quantum effects, limiting the observed quantum noise squeezing to approximately 2-3 dB in most integrated platforms \cite{zhao2020near,yang2021squeezed,liu2025wafer} and thus restricting the performance of prospective on-chip quantum sensors and quantum computers.

A straightforward solution to this issue of coupling losses is to integrate optical detectors on-chip as well, but the challenge is formidable due to difficulties in heterogeneous integration of high-performance materials (such as germanium or III-V compounds) into a standard CMOS foundry process \cite{najafi2015chip}. On the other hand, recent theoretical and experimental advances \cite{shaked2018lifting,li2019pulsed,li2020measuring,kashiwazaki2021fabrication} have led to a new paradigm that can circumvent this problem. Instead of trying to minimize losses after the squeezer, this approach uses an optical parametric amplifier (OPA) as a part of a loss-resilient measurement device. The OPA amplifies the input signal without adding extra noise before detection, elevating its noise floor far above vacuum noise that would otherwise enter through lossy channels. This strategy was recently applied to the detection of pulsed squeezing on a thin-film PPLN chip \cite{nehra2022few}.

In fact, a particularly powerful realization of this concept was demonstrated in the SU(1,1) quantum interferometer \cite{yurke19862,jing2011realization,hudelist2014quantum,ou2020quantum}, which consists of two OPAs for the replacement of traditional linear beam splitters in classical interferometers. The first OPA generates a squeezed/entangled state, whereas the second OPA amplifies and measures it. This arrangement preserves quantum correlations, leading to enhancement in signal-to-noise ratio, and thus improves the sensitivity of phase sensing in this quantum interferometer. The quantum enhancement effect in this interferometer has been shown to be tolerant to external losses such as propagation and detection inefficiency\cite{hudelist2014quantum,ou2012enhancement,manceau2017detection} due to the use of the second parametric amplifier.

In this work, we report the experimental realization of this paradigm on a silicon nitride (SiN) integrated photonic platform with two cascaded micro-ring resonators in which a two-mode squeezed state of light is generated in the first ring via a $\chi^{(3)}$-nonlinear process and subsequently measured in situ by a high-gain parametric amplifier in the second ring before the light is subject to significant off-chip coupling losses before detection. 4.6 dB of quantum noise reduction is observed at the output of the device, even with 7 dB of overall system losses from chip to detection. This is a substantial advance for integrated quantum optics and a demonstration of loss-resilient quantum measurement on a chip.

Our device also constitutes a fully integrated SU(1,1) interferometer, where the generation and interference of quantum-correlated photons occur within a single, compact circuit. This work fundamentally reorients the strategy for managing loss in integrated quantum photonics, moving from the mitigation of external loss to its operational circumvention. Hence, it establishes a practical and scalable pathway for chip-based quantum sensing, metrology, and information processing, where the preservation of quantum properties is paramount.


\section*{Results}
\subsection*{Principle of loss-tolerant, amplifier-assisted measurement}

The idea of using amplifiers to circumvent detection losses was first proposed by Caves in his seminal work on quantum noise in interferometers \cite{cav81}. A similar strategy was later employed to overcome substantial thermal noise in squeezing measurements and in the demonstration of Einstein--Podolsky--Rosen (EPR) correlations in the microwave domain \cite{mallet11,flurin12}. More recently, this approach has been extended to the optical regime, with implementations in free-space environments using pulsed  \cite{shaked2018lifting,li2019pulsed,che21,nehra2022few} and continuous-wave (CW) operation \cite{kashiwazaki2021fabrication}. 

\begin{figure} 
	\centering
	\includegraphics[width=1.0\textwidth]{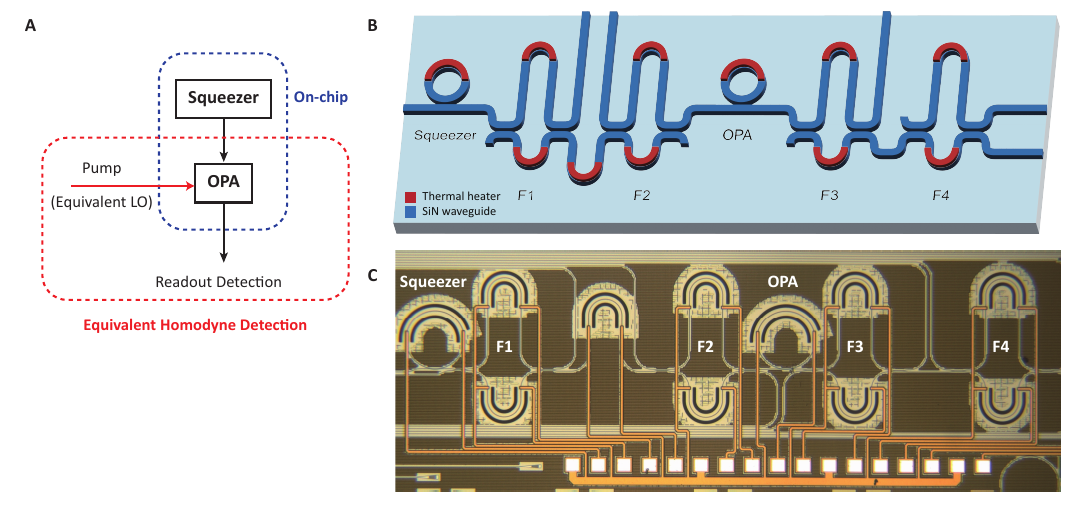}
    \caption{\textbf{Concept and device architecture of on-chip parametric-amplifier-assisted quantum measurement.}
		  (\textbf{A}) Conceptual schematic of parametric-amplifier-assisted quantum measurement. A high-gain optical parametric amplifier (OPA) measures the squeezed light, with its pump serving as the equivalent local oscillator (LO). The blue and red dashed regions indicate the on-chip section and equivalent homodyne detection stage, respectively.
        (\textbf{B}) Schematic diagram of the on-chip device structure, comprising two ring-resonators, denoted as the Squeezer and OPA, and four cascaded MZI filters (F1-F4).
        (\textbf{C}) Device photograph. }\label{fig:device}
\end{figure}

The conceptual schematic is shown in Fig.~\ref{fig:device}(A). The output of a quantum source is coupled into a high-gain optical parametric amplifier (OPA) prior to detection. Depending on the quantum source, whose output can be in single mode or in two modes, we can use a degenerate or non-degenerate parametric amplifier \cite{li2020measuring}. For our situation here, where a two-mode squeezed state is generated from a micro-ring with $\chi^{(3)}$ nonlinearity through a four-wave mixing process, we need a non-degenerate parametric amplifier, whose output fields are related to the input fields as \cite{li2020measuring}:
\begin{eqnarray}
\hat{a}_{ {\text{out }}}^{s,i} =  G \hat{a}_{\text{in}}^{s,i} + g e^{2i\phi_p} \hat{a}_{\text{in}}^{i,s \dag} , \label{PA}
\end{eqnarray}
where ``s, i" denote the two modes commonly referred to as ``signal, idler" in parametric processes, and $G,g$ are the amplitude gain parameters satisfying $G^2-g^2=1$. $\phi_p$ is the phase of the pump field to the amplifier using four-wave mixing as the parametric process. Concentrating on the signal output only, its quadrature-phase amplitude $\hat X(\theta) \equiv \hat a e^{-i\theta} +\hat a^{\dag}e^{i\theta}$ becomes
\begin{eqnarray}
\hat{X}_{ {out }}^{s}(\theta) &=&  G \hat{X}_{in}^{s}(\theta) + g \hat{X}_{in}^{i}(2\phi_p-\theta)\cr  &=& G [\hat{X}_{in}^{s}(\theta) + \lambda \hat{X}_{in}^{i}(2\phi_p-\theta)] \cr &\rightarrow & G \hat X_+(\theta,\phi_p) ~~~{\rm for}~~G\gg 1, \label{X}
\end{eqnarray}
where $\lambda \equiv g/G \rightarrow 1$ for $G \gg 1$ and $\hat X_+(\theta,\phi_p)\equiv \hat{X}_{in}^{s}(\theta) + \hat{X}_{in}^{i}(2\phi_p-\theta)$. 

For input fields in a two-mode squeezed state, the quantum correlation between the input signal and idler fields will lead to quantum noise reduction (squeezing) in $\hat X_+(\theta,\phi_p)$ at the appropriate phases of $\theta,\phi_p$ compared to the corresponding uncorrelated vacuum noise level of $\langle \Delta^2\hat X_+(\theta,\phi_p)\rangle_{vac} = 2$. From Eq.~(\ref{X}), we can find the output noise level of the amplifier in the large gain limit as
\begin{eqnarray}
\langle \Delta^2 \hat{X}_{ {out }}^{s}(\theta) \rangle &=& G^2 \langle \Delta^2\hat X_+(\theta,\phi_p)\rangle ~~~{\rm for}~~G\gg 1. \label{N}
\end{eqnarray}
Therefore, comparing the output noise for a squeezed input (``sq'') versus a vacuum input (``vac'') yields the noise reduction factor:
\begin{eqnarray}
{\cal R}\equiv \frac{\langle \Delta^2 \hat{X}_{ {out }}^{s}(\theta) \rangle_{sq}}{\langle \Delta^2 \hat{X}_{ {out }}^{s}(\theta) \rangle_{vac}} = \frac{\langle \Delta^2\hat X_+(\theta,\phi_p)\rangle_{sq}}{\langle \Delta^2\hat X_+(\theta,\phi_p)\rangle_{vac}}. \label{R}
\end{eqnarray}
So, by measuring the output noise level of the amplifier with squeezed state input and comparing it with the case when the input is in vacuum, the noise reduction in the output of the amplifier directly gives the amount of squeezing in the input of the amplifier as compared to the vacuum noise. 

Next, let us show that this scheme is tolerant to losses in readout detection. This is because of the high output noise level of the amplifier compared to vacuum noise, which comes through losses. Actually, the detected noise is given by
\begin{eqnarray}\label{L}
\langle \Delta^2 \hat X _{out}^{det} \rangle = (1-L)\langle \Delta^2 \hat X _{out} \rangle +L 
\end{eqnarray}
with $L$ as the overall loss before ideal detection (100\% quantum efficiency). When the inputs are in a two-mode squeezed state with a noise reduction factor of $\cal R$, the output noise level (Eq.~(\ref{N})) is $\langle \Delta^2 \hat X _{out} \rangle \approx  2G^2 {\cal R} $ for $G\gg1$ according to Eq.~(\ref{R}). So, as long as $(1-L) 2G^2 {\cal R} \gg 1$ for large enough $G^2$, the added vacuum noise ($L$) through the lossy channel is negligible even for a large loss $L\sim 1$. Combining Eqs.(\ref{R}, \ref{L}) and neglecting the vacuum contribution in the last term in Eq.~(\ref{L}), we have the noise reduction factor at the detector 
\begin{eqnarray}
{\cal R}^{det} &\equiv& \frac{\langle \Delta^2 \hat{X}_{ {out }}^{det}(\theta) \rangle_{sq}}{\langle \Delta^2 \hat{X}_{ {out }}^{det}(\theta) \rangle_{vac}} \cr &\approx & \frac{(1-L)G^2\langle \Delta^2\hat X_+(\theta,\phi_p)\rangle_{sq}}{(1-L)G^2\langle \Delta^2\hat X_+(\theta,\phi_p)\rangle_{vac}} ={\cal R}. \label{R-d}
\end{eqnarray}
So, the loss has no effect on the detected noise reduction.  

Notice that Eq.~(\ref{X}), in the limit of large gain, shows noiseless amplification that does not add extra noise as in a regular quantum amplifier. So, the loss tolerance of the amplification scheme is because the delicate quantum noise is first amplified noiselessly, before any loss, to a large classical noise level that is immune to losses. This property enables direct on-chip measurement of squeezing before off-chip coupling losses can destroy it.

Another perspective is that the pump functions analogously to a strong local oscillator (LO) in conventional homodyne detection, where the LO amplifies the optical signal to surpass electronic noise. However, in the case here,  the pump's role is to amplify the quantum correlation to overcome vacuum noise introduced through losses (Fig.~\ref{fig:device}(A)). So, we can think of it as equivalent to the quantum measurement performed on-chip.


\subsection*{Device structure and linear characterization}


Both the squeezer and the OPA used in this work are implemented on a silicon nitride (SiN) integrated photonic platform and are based on a $\chi^{(3)}$ nonlinear four-wave mixing process in micro-ring resonators. Silicon nitride has emerged as a versatile and high-performance platform for integrated nonlinear photonics, offering high refractive-index contrast, low propagation loss, and a broad transparency window from visible to mid-infrared~\cite{cui2023compact,moss2013new,blumenthal2018silicon}. 
Specifically, when pumped above threshold, these devices operate as optical parametric oscillators (OPOs), which can initiate cascaded four-wave mixing and lead to Kerr frequency combs, providing a route to broadband chip-scale light sources~\cite{gaeta2019photonic,levy2010cmos,okawachi2011octave}.
In parallel, when pumped far below the oscillation threshold, SiN micro-rings can generate entangled photon pairs at telecom wavelengths through spontaneous four-wave mixing (SFWM), providing a compact and CMOS-compatible platform for integrated quantum information processing and quantum computing~\cite{kues2019quantum,samara2019high,imany2018fifty}.
In this work, we operate these devices just below the threshold, where they function as high-gain OPAs, enabling the generation of two-mode squeezed states~\cite{zhang2021squeezed,yang2021squeezed,liu2025wafer} and the on-chip measurement of squeezing.

The schematic of the on-chip structure is shown in Fig.~\ref{fig:device}(B), and the chip micrograph is shown in Fig.~\ref{fig:device}(C). 
Two micro-ring resonators are placed in series: the first (R1, Squeezer) generates a two-mode squeezed state in the signal and idler modes via SFWM by a strong on-resonance pump, while the second (R2, OPA) takes R1's output as the input and acts as a phase-sensitive parametric amplifier for measuring squeezing from R1.
Between the two rings, two cascaded unbalanced Mach--Zehnder interferometer (MZI) filters (F1, F2) are monolithically integrated to effectively filter out the pump while ensuring the passing of the signal and idler generated in adjacent resonances. Thermal heaters are attached to each of these devices individually for independent fine temperature tuning. The two coupled micro-ring resonators also form an on-chip SU(1,1) interferometer capable of quantum sensing through precision phase measurement. 

The fabrication of the microchip is done on an 800-nm-thick silicon nitride wafer via a multi-project wafer shuttle run by a commercial foundry (Ligentec). More specifically, two size-matched micro-ring resonators are designed with a radius R = 114 $\mu m$, yielding free spectral ranges (FSRs) of approximately 200 GHz. The length difference between the two arms of the unbalanced MZI filters (F1,F2) is designed to be 358.58 $\mu m$, resulting in an FSR of approximately 400 GHz (twice the resonators' FSR). The two rings are resonantly pumped, respectively, at 1557 nm. Through four-wave mixing (FWM), signal and idler fields are generated resonantly in both rings, with maximum gains occurring at 1542 nm and 1572 nm, respectively.  The pump to the first ring is filtered out by F1 while the pump to the second ring is coupled in through F2. These two filters provide $>$ 30 dB extinction for the pump fields while allowing the signal and idler fields to pass with minimum loss, ensuring efficient coupling between the two rings. Additional filters (F3, F4) channel and filter out the pump to the second ring. The ring resonator is designed with a width of 1.7 $\mu m$, and a bend directional coupler is used to achieve over-coupling in the coupling region. 

\begin{figure}[htbp]
  \centering
\includegraphics[width=1.0\textwidth]{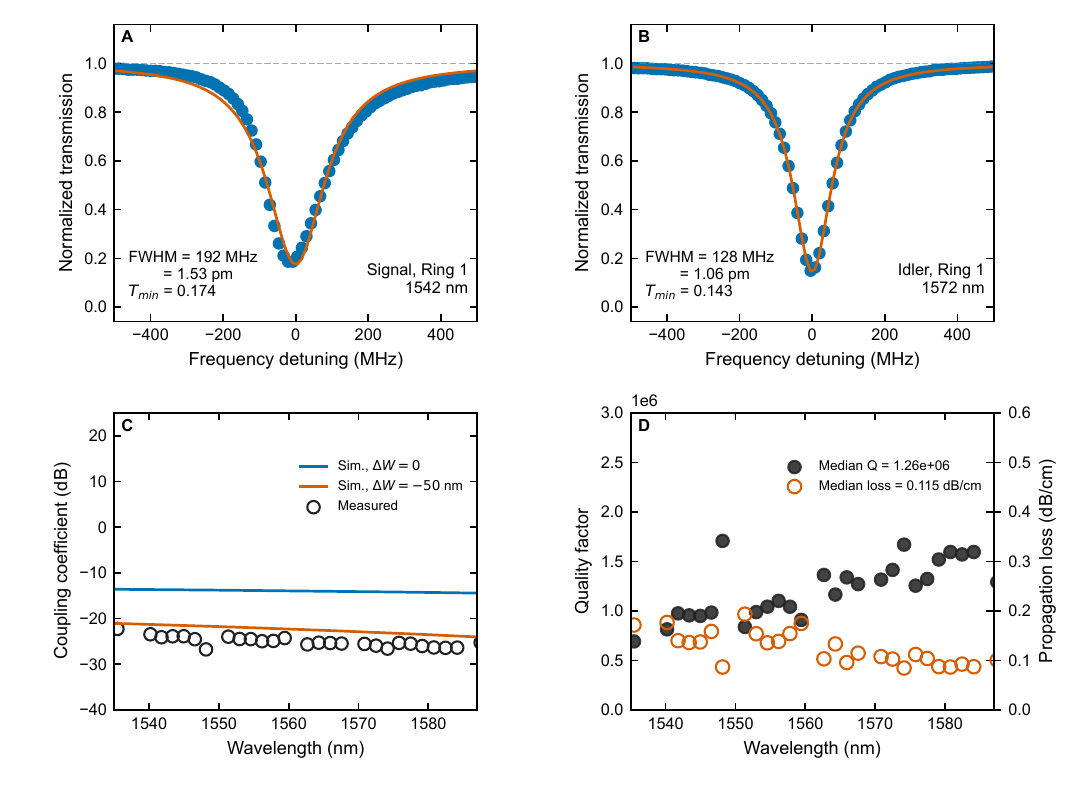}
  \caption{\textbf{Ring linear characterization.}
    (\textbf{A})(\textbf{B}) Measured transmission spectrum at signal wavelength of 1542~nm and idler wavelength of 1572~nm, respectively. 
    (\textbf{C}) Simulated and measured coupling coefficient $\kappa$.
    (\textbf{D}) Calculated loaded Q factor and waveguide propagation loss $\alpha$.}
  \label{fig:linear}
\end{figure}

To characterize the linear properties of R1, we measure its transmission spectra by injecting a tunable laser into the input port of the chip.
The measured transmission spectra for the first ring are shown in Fig.~\ref{fig:linear}(A, B) at the signal and idler wavelengths, respectively.
The coupling coefficients $\kappa$, the loaded Q factors, and the propagation loss $\alpha$ were then extracted from the measured transmission and are shown in Fig.~\ref{fig:linear}(C, D).
High squeezing level requires an over-coupled design~\cite{zhao2020near,vaidya2020broadband}.
We target a coupling ratio of approximately 5\%. However, due to fabrication variations (a 50 nm reduction in waveguide width), the measured coupling ratio $\kappa$ is approximately 0.3\% (see Fig.~\ref{fig:linear}(C)). The measured propagation loss is approximately 0.11 dB/cm, corresponding to an intrinsic Q over 3 million (M) and the loaded Q is approximately 1.26 M, as shown in Fig.~\ref{fig:linear}(D). The measured extinction ratios in Fig.~\ref{fig:linear}(A, B) yield estimated escape efficiencies of approximately $\eta_{\text{esp}}\equiv\kappa/(\alpha+\kappa)\approx$ 71\% and 69\% for the signal and idler field, respectively, assuming the over-coupled branch. 
The estimated escape efficiencies are used to evaluate the quantum noise reduction at the output of the squeezer, with a detailed discussion provided in the Discussion section.


\subsection*{Gain characterization of OPAs}
\begin{figure} 
	\centering
	\includegraphics[width=1.0\textwidth]{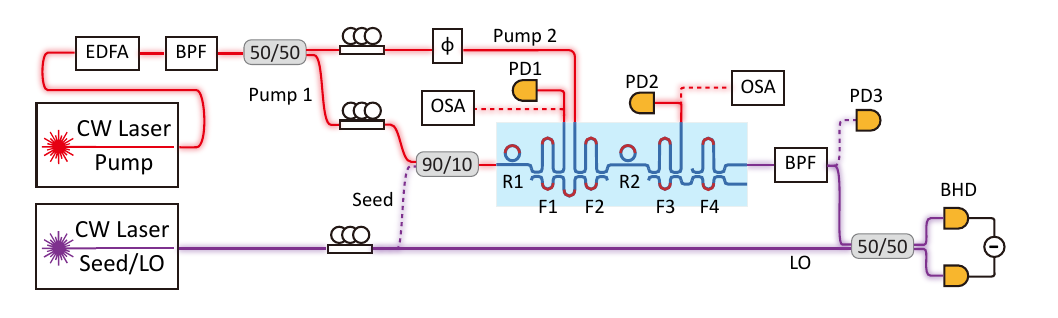} 
	\caption{\textbf{Experimental setup.} The blue box encloses the on-chip device, which consists of two ring-resonator-based OPAs (R1 and R2) and cascaded unbalanced Mach-Zehnder interferometer filters (F1-F4). For quantum-noise squeezing measurements, only the pump is injected. For classical interference fringe measurement, an additional seed field is injected into the device to characterize the phase-dependent response of the cascaded OPAs. EDFA, erbium-doped fiber amplifier; BPF, bandpass filter; BHD, balanced homodyne detector; OSA, optical spectrum analyzer.}
	\label{fig:setup} 
\end{figure}
\begin{figure}[htbp]
    \centering
    \includegraphics[width=0.6\textwidth]{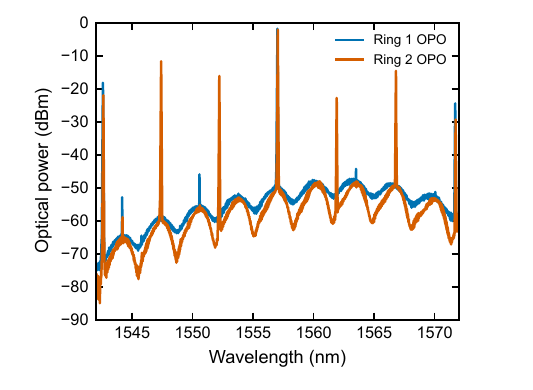}
    \caption{\textbf{OPO characterization.} Overlap between the OPO spectra of the two micro-ring resonators, with matched signal and idler modes at 1542~nm and 1572~nm, respectively.}
    \label{fig:OPO}
\end{figure}

We characterize the nonlinear properties of the two ring resonators by pumping them separately to reach the above-threshold optical parametric oscillation (OPO) regime and monitoring the output spectra with an optical spectrum analyzer (OSA). The experimental setup is shown in Fig.~ \ref{fig:setup}.  Light is coupled into and out of the chip via edge coupling to a fiber array (FA).  Pump 1 is directly coupled into the first ring (R1), through the chip input port, then filtered out by F1 and monitored by PD1. Pump 2 is coupled into the second ring (R2) through the F2 input, filtered out by F3, and monitored by PD2. We tune the temperatures of R1 and R2 separately by controlling the voltages applied to the thermal heaters attached to them. This enables us to place the resonances of both rings as close as possible, which is critical for efficient coupling between the two rings. The pump wavelength is also tuned to achieve resonance for optimum nonlinear interaction inside the micro-rings.  The OPO spectra of R1 and R2 are measured at the output ports of F1 and F3, respectively, using an OSA and are displayed together in Fig.~\ref{fig:OPO}, which shows the mode structure with overlap in both the 1542~nm and 1572~nm sidebands.

\begin{figure}[htbp]
    \centering
    \includegraphics[width=0.6\textwidth]{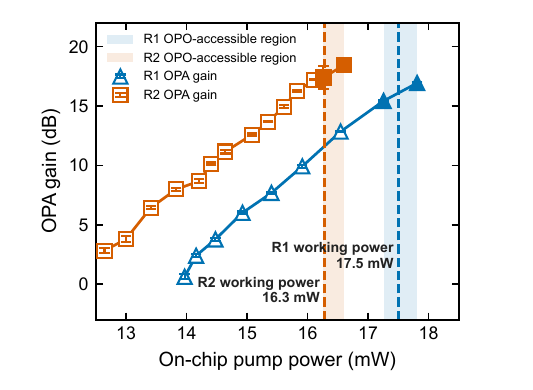}
    \caption{\textbf{OPA gain characterization.} Measured OPA gain versus on-chip pump power for R1 and R2. Open and filled markers represent the OPA and OPO regimes, respectively. Shaded areas indicate the OPO-accessible regions, and vertical dashed lines mark the experimental working points.}
  \label{fig:opa_gain}
\end{figure}

Squeezing generation and amplifier-assisted quantum measurement require nonlinear devices operating just below the threshold to have high gains~\cite{zhao2020near,vaidya2020broadband}. 
The measured OPO threshold powers are approximately 
$17.3~\mathrm{mW}$ and $16.3~\mathrm{mW}$
on-chip for R1 and R2, respectively (see the first filled marker on each trace in Fig.~\ref{fig:opa_gain}). The corresponding working pump powers are $17.5~\mathrm{mW}$ for R1 and $16.3~\mathrm{mW}$ for R2 respectively. 
Although these pump powers are very close to the respective OPO thresholds, the pumps are slightly detuned from the peak-gain resonances and remain below threshold. This choice is important because thermal effects and cross-phase modulation jointly induce cavity bistability, making the maximum-gain point sensitive to small pump-frequency fluctuations. We therefore operate at points that provide appreciable and stable parametric gain while reducing the risk of leaving the high-gain regime.

In addition, we measured the classical gains of the devices with seed injection, as shown in Fig.~\ref{fig:setup}. The measured gains of the corresponding OPAs are plotted in Fig.~\ref{fig:opa_gain} as a function of on-chip pump power, reaching values as high as 17~dB. The actual gains are expected to be higher, as the measured gains include losses inside the rings.

\subsection*{Quantum noise measurement and observation of squeezing}
\begin{figure}[htbp]
    \centering
    \includegraphics[width=1.0\textwidth]{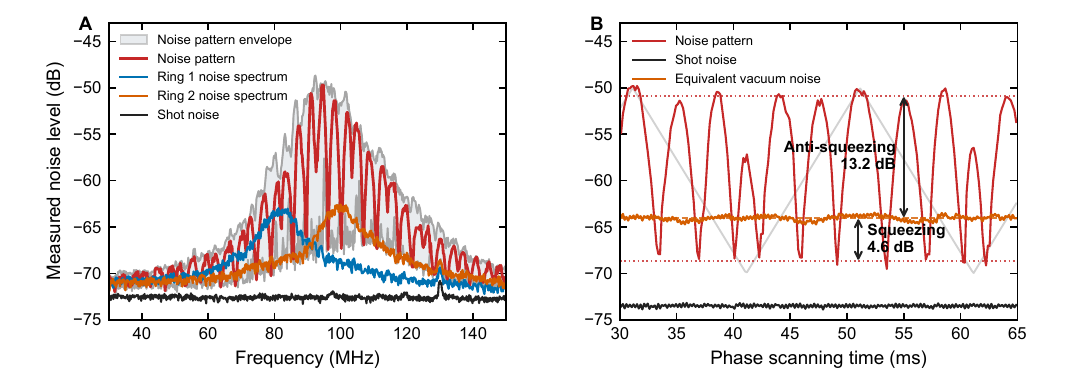}
    \caption{\textbf{Quantum noise measurement.} 
        (\textbf{A}) Quantum-noise spectra measured by an electronic spectrum analyzer (ESA), showing the noise level versus analysis frequency. The blue and orange traces show the individual noise spectra of R1 and R2, respectively, with the other ring turned off. The black trace indicates the vacuum-noise level (shot noise). The red trace is measured with both rings pumped while scanning the phase of Pump 2. The gray curves indicate the noise pattern minima and maxima extracted from multiple averaged phase scans.
        (\textbf{B}) Noise levels measured in zero-span mode with the ESA center frequency set to 100~MHz, where R2 is operated at maximum gain.  The orange trace shows the noise level of R2 alone, measured with only Pump~2 on and Pump~1 to R1 turned off, providing the equivalent vacuum-noise reference for the output of R1. The red trace shows the noise level as a function of the Pump~2 phase scan when both R1 and R2 are pumped. Dashed lines indicate the average values of the corresponding noise levels. The gray curve shows the ramp voltage used for the phase scan.}
    \label{QN}
\end{figure}

The quantum noise of the system was measured using a balanced homodyne detector (BHD) at the chip output, with the seed blocked and only the pump injected, as shown in Fig.~\ref{fig:setup}.
According to the theory presented above, it is sufficient to measure only the signal field. We therefore derive the 1~mW local oscillator (LO) for the BHD from the seed laser at 1542~nm, which matches the signal wavelength.
Based on Eq.~(\ref{R}), the output noise level of the second ring with vacuum input provides the equivalent vacuum-noise reference for comparison, while the amount of noise squeezing is determined by the gain of the first ring. We therefore first measured the quantum-noise level of each ring separately, with the other ring turned off. Because the BHD measured only the signal field using an LO at 1542~nm, the measured noise level corresponds only to the quantum noise gain in the signal field.
The quantum-noise spectra measured by the electronic spectrum analyzer (ESA) are summarized in Fig.~\ref{QN}(A). The black trace provides the vacuum-noise reference when the input to BHD is blocked. The single-ring measurements (blue and orange traces) show that the two rings exhibit slightly different gain profiles, mainly due to a mismatch between their free spectral ranges. When both pumps are applied and the phase of pump 2 is scanned, the measured noise (red trace) becomes phase sensitive, showing clear noise amplification and de-amplification. From these phase-dependent spectra, we extract the maximum and minimum noise levels as a function of the analysis frequency, shown as the upper and lower envelopes in Fig.~\ref{QN}(A).


Notice that the gain spectral profiles are centered around 100 MHz. This is because both the seed and pump lasers are free running, and the measurement of only the signal field does not require the LO laser to be locked on resonance. 
To measure maximum squeezing, we use the zero-span mode with the ESA center frequency set to 100~MHz, where the quantum noise level of R2 reached its maximum. This choice was made because a large gain in the measurement OPA is required for a faithful squeezing measurement. The resolution bandwidth (RBW) and video bandwidth (VBW) were set to 1~MHz and 1~kHz, respectively.
The zero-span measurements are summarized in Fig.~\ref{QN}(B). We first established the shot-noise reference (black trace) by blocking the BHD input. We then measured the noise level of R2 alone, with Pump~1 to R1 turned off, providing the equivalent vacuum-noise reference for the output of R1 (orange trace). With both rings pumped, the quantum-correlated signal and idler fields generated in R1 were coupled into R2 for on-chip quantum measurement. The result is shown as the red trace in Fig.~\ref{QN}(B) when the phase $\phi$ of Pump~2 is scanned. Relative to the equivalent vacuum-noise reference, the minimum noise level was reduced by 4.6 $\pm$ 0.4~dB, corresponding to the measured squeezing, while the maximum noise level increased by 13.2 $\pm$ 0.4~dB, corresponding to the measured anti-squeezing.
This measured 4.6 dB noise reduction directly corresponds to the quantum correlation between the signal and idler fields or the two-mode squeezing level generated by the first ring, as discussed in Eq.~(\ref{R}).

We further calibrate the optical losses of the system: the signal and idler are coupled off chip with a measured coupling loss of 5 dB, and only the signal is retained using a bandpass filter (BPF) with an insertion loss of about 1 dB. From the manufacturer's data sheet, we find that the quantum efficiency of the BHD detectors is about 80\% (1 dB). Therefore, the total estimated downstream loss after R2 is approximately $7~\mathrm{dB}$. 


Therefore, a quantum noise reduction of 4.6~dB was observed despite a measured overall loss (from chip to detection) of approximately 7~dB, which would have reduced the observable squeezing to less than 1~dB under conventional direct homodyne detection without the aid of the amplifier. We also note that the measured noise level of R2 alone (the orange trace in Fig.~\ref{QN}(B)) is about 10~dB above the shot-noise level. Taking into account the overall loss of approximately 7~dB, the corresponding on-chip quantum-noise gain of R2 is estimated to be about 17~dB. This confirms that R2 operates as a high-gain parametric amplifier, as required for on-chip quantum measurement. Similarly, the measured noise level of R1 alone (not shown) is 5~dB above the shot-noise level, corresponding to an on-chip amplified quantum-vacuum-noise level of approximately 12~dB after correcting for the same loss.

\subsection*{On-chip SU(1,1) Interferometer with Injected Seed}
\begin{figure}[htbp]
  \centering
  \includegraphics[width=0.6\textwidth]{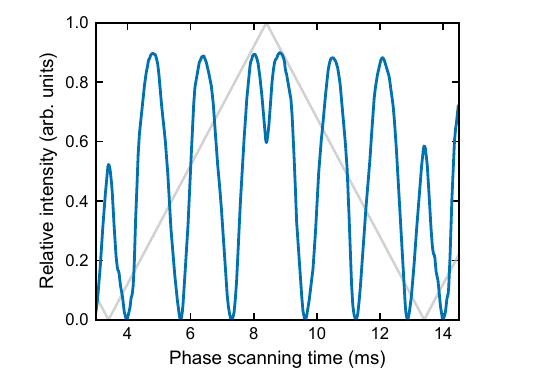
  }
  \caption{\textbf{Interference with injected seed.} Interference fringe of the SU(1,1) interferometer formed by Ring 1 and 2 when a seed is injected in the signal field. The gray curve shows the ramp voltage for phase scan.}
  \label{SUI}
\end{figure}

The experimental arrangement with an injected seed is also a realization of an on-chip SU(1,1) interferometer, where R1 and R2, serving as non-degenerate optical parametric amplifiers (OPA1, OPA2), act as equivalent beam splitters in replacement of linear beam splitters of conventional interferometers \cite{ou2020quantum}: the injected seed in the signal field is first split into amplified signal and companion idler fields by OPA1, which are then recombined by OPA2 to form an interferometer for interference.  
With a weak seed signal injected into R1 (OPA1), we observe the intensity at the output port of the chip  by PD3 via a band pass filter (BPF) at the signal wavelength as the phase of Pump 2 is scanned. Figure~\ref{SUI} shows the resulting interference fringe with a visibility of 98.0\%. The clear fringe demonstrates that the cascaded rings function as a monolithic SU(1,1) interferometer, capable of phase-sensitive measurement with potential sensitivity beyond the shot-noise limit \cite{ou2020quantum}. 

SU(1,1) interferometers are known to provide quantum enhancement of phase measurement sensitivity compared to traditional Mach-Zehnder interferometer \cite{ou2012enhancement}. Although we cannot directly measure the quantum enhancement due to limitation in chip design (phase scanning is only done by Pump 2), we can estimate it \cite{hudelist2014quantum} using the available experimental data measured so far. The working principle of quantum enhancement of SU(1,1) interferometers is the amplification of the phase signal by the second parametric amplifier while quantum noise is reduced due to destructive interference. The classical gain of R2 provides the amplification of the phase signal produced inside the interferometer (between the two rings) compared to a traditional linear interferometer whose signal is not amplified. From the gain data of OPA2 in Fig.~\ref{fig:opa_gain}, we find that the signal gain from OPA2 is around 17 dB. The observed phase signal gain will be 7 dB less due to losses, which gives $17-7$ = 10 dB of the detected phase signal gain compared to a traditional linear interferometer. With respect to noise, a traditional linear interferometer is known to operate at the vacuum noise level (shot noise). From Fig.~\ref{QN}(B), we find that the output quantum noise (the average of the minima of the red trace in Fig.~\ref{QN}(B)) of the on-chip SU(1,1) interferometer is about 5 dB above the shot noise level (vacuum noise level). Therefore, the quantum enhancement of the signal-to-noise ratio (SNR) is around $10 - 5 \approx 5$ dB, which is consistent with the observed quantum noise squeezing of 4.6 dB.

\section*{Discussion}

In summary, we demonstrate an on-chip quantum measurement scheme with the aid of a parametric amplifier. Using such a scheme, we observed 4.6 dB of quantum noise squeezing generated from an on-chip squeezer, even with the presence of a large loss of more than 7 dB from the chip to the detection mainly due to chip coupling loss. Compared to other on-chip measurement schemes such as on-chip integration of detector, our approach is all optical, and since the optical parametric amplifier is part of the on-chip nonlinear devices, it is easier to fabricate. 
Our scheme is also a genuine quantum interferometer with a quantum enhancement of phase measurement sensitivity with an indirectly estimated quantum enhancement factor of about 5 dB. 

Due to limitations in chip design and fabrication errors, the escape efficiencies $\eta_{esp}$ of the squeezing generation micro-ring are only 71\% at the signal and 69\% at the idler wavelengths, or an average of $\bar\eta_{esp} =70\%$, which would yield a maximum squeezing of $1- \bar\eta_{esp} = 0.3 = - 5.2 dB$ on average out of the squeezer.  This sets the upper limit to the observed squeezing, which is $-4.6$ dB here. The discrepancy may come from the coupling loss between the two rings and the large detuning in the squeezer ring.  Our next batch of chip design will have a run of parameters to ensure a higher escape efficiency for a larger amount of squeezing from the squeezer. 

At this moment, the phase scan of the SU(1,1) interferometer is carried out externally through Pump 2. So, the quantum enhancement of the phase measurement sensitivity can only be estimated on the basis of other measured information. To directly measure the quantum enhancement, we need to have the phase modulation signal on-chip and make a comparison with an on-chip linear interferometer.  The pump splitting will be performed on-chip as well to increase the phase control stability of the interferometer. These will be implemented in the next batch of chip design.

 In conclusion, our work will enable the development of highly compact and portable quantum sources and establish a pathway towards chip-scale quantum sensors. The amplifier-assisted on-chip quantum measurement scheme is expected to play a crucial role in the future of integrated quantum optics, with particular relevance to chip-based optical quantum computing~\cite{Clark_2026_Integrated,Gonzalez-Arciniegas_2021_Cluster} and quantum sensing applications~\cite{Aasi_2013_Enhanceda,Herman_2025_Squeezed}.

\clearpage

\bibliography{bibliography}
\bibliographystyle{sciencemag}


\section*{Acknowledgments}

\paragraph*{Funding:}
This work is supported by City University of Hong Kong (Project No. 9610522), the General Research Fund of the Hong Kong Research Grants Council (No. 11307823), and the State Key Laboratory of Quantum Information Technologies and Materials (The Chinese University of Hong Kong).
\paragraph*{Author contributions:}
Z. Y. Ou conceived the idea. C. C. designed the chip. Y. Lei carried out the experiments and analyzed the data with the help from C. C., Y. Li, and Z. Y. Ou. Y. Lei and C. C. performed theoretical calculations, supervised by Z. Y. Ou and H. K. Tsang. Y. Lei, C. C. and Z. Y. Ou wrote the manuscript with contributions from all authors. Z. Y. Ou and H. K. Tsang supervised this project.
\paragraph*{Competing interests:}
There are no competing interests to declare.
\paragraph*{Data and materials availability:}
All data needed to evaluate the conclusions in the paper are present in the paper.












\end{document}